\documentclass[onecolumn,11pt,msmath,superscriptaddress,amsmath,amssymb,aps]{revtex4-1}
\usepackage[colorlinks=true,urlcolor=blue,linkcolor=blue,citecolor=blue]{hyperref}
\usepackage{subfigure}
\usepackage{tablefootnote}
\usepackage{wrapfig}
\usepackage{graphicx}
\usepackage{textcomp}
\usepackage{gensymb}
\usepackage{bm}
\usepackage{xfrac}
\usepackage{color}
\usepackage{xcolor}
\usepackage{setspace}
\usepackage{silence}
\usepackage{epstopdf}
\RequirePackage[normalem]{ulem} 
\RequirePackage{color}
\definecolor{RED}{rgb}{1,0,0}
\definecolor{BLUE}{rgb}{0,0,1}

\makeatletter
    \renewcommand\@make@capt@title[2]{%
     \@ifx@empty\float@link{\@firstofone}{\expandafter\href\expandafter{\float@link}}%
      {\textbf{#1}}\@caption@fignum@sep#2\quad}%
    \makeatother

\usepackage{array}
\newcolumntype{C}[1]{>{\centering\arraybackslash}m{#1}}
\usepackage[final]{changes}

\begin{document}
\title{Fermi surface and kink structures in Sr$_{4}$Ru$_{3}$O$_{10}$ revealed by synchrotron-based ARPES}
\author{Prosper Ngabonziza}
\email{p.ngabonziza@fkf.mpg.de}
\affiliation{Max Planck Institute for Solid State Research, Heisenbergstraße 1, 70569 Stuttgart, Germany}
\affiliation{Department of Physics, University of Johannesburg, PO Box 524, Auckland Park 2006, South Africa}

\author{Emanuela Carleschi}
\email{ecarleschi@uj.ac.za}
\affiliation{Department of Physics, University of Johannesburg, PO Box 524, Auckland Park 2006, South Africa}
\author{Volodymyr Zabolotnyy}
\affiliation{Physikalisches Institut, Julius-Maximilians-Universität Würzburg, \\
Am Hubland,97074 Würzburg, Germany}
\author{Amina Taleb-Ibrahimi}
\affiliation{Synchrotron SOLEIL, L’Orme des Merisiers, Saint-Aubin-BP48, 91192 Gif-sur-Yvette, France}
\author{François  Bertran}
\affiliation{Synchrotron SOLEIL, L’Orme des Merisiers, Saint-Aubin-BP48, 91192 Gif-sur-Yvette, France}
\author{Rosalba Fittipaldi}
\affiliation{CNR-SPIN Salerno, Via Giovanni Paolo II, 84084 Fisciano, Italy}
\affiliation{Department of Physics, University of Salerno, Via Giovanni Paolo II, 84084 Fisciano, Italy}
\author{Veronica Granata}
\affiliation{CNR-SPIN Salerno, Via Giovanni Paolo II, 84084 Fisciano, Italy}
\affiliation{Department of Physics, University of Salerno, Via Giovanni Paolo II, 84084 Fisciano, Italy}

\author{Mario Cuoco}
\affiliation{CNR-SPIN Salerno, Via Giovanni Paolo II, 84084 Fisciano, Italy}
\affiliation{Department of Physics, University of Salerno, Via Giovanni Paolo II, 84084 Fisciano, Italy}

\author{Antonio Vecchione}
\affiliation{CNR-SPIN Salerno, Via Giovanni Paolo II, 84084 Fisciano, Italy}
\affiliation{Department of Physics, University of Salerno, Via Giovanni Paolo II, 84084 Fisciano, Italy}
\author{Bryan Patrick Doyle}
\email{bpdoyle@uj.ac.za}
\affiliation{Department of Physics, University of Johannesburg, PO Box 524, Auckland Park 2006, South Africa}

\begin{abstract}

\section*{\Large{A\lowercase{bstract}}}
The low-energy electronic structure, including the Fermi surface topology, of the itinerant metamagnet Sr$_{4}$Ru$_{3}$O$_{10}$ is investigated for the first time by synchrotron-based angle-resolved photoemission. Well-defined quasiparticle band dispersions with matrix element dependencies on photon energy or photon polarization are presented. Four bands crossing the Fermi-level, giving rise to four Fermi surface sheets are resolved; and their complete topography, effective mass as well as their electron and hole character are determined. These data reveal the presence of kink structures in the near-Fermi-level band dispersion, with energies ranging from 30 meV to 69 meV. Together with previously reported Raman spectroscopy and lattice dynamic calculation studies, the data suggest that these kinks originate from strong electron-phonon coupling present in Sr$_{4}$Ru$_{3}$O$_{10}$. Considering that the kink structures of Sr$_{4}$Ru$_{3}$O$_{10}$  are similar to those of the other three members of the Ruddlesden Popper structured ruthenates, the possible universality of strong coupling of electrons to oxygen-related phonons in Sr$_{n+1}$Ru$_{n}$O$_{3n+1}$ compounds is proposed. 
\end{abstract}

\pacs{???}
\maketitle

\section*{\Large{I\lowercase{ntroduction}}}
\paragraph*{}Strontium ruthenate oxides of the Ruddlesden Popper (RP) series,  Sr$_{n+1}$Ru$_{n}$O$_{3n+1}\, (n= 1, 2, 3, \infty)$, are a class of materials in the realm of  ${4d}$ transition metal oxides that are attractive for both fundamental and applied research. From the perspective of fundamental physics, Sr$_{n+1}$Ru$_{n}$O$_{3n+1}$ are strongly correlated materials which exhibit complex interplay between the charge, spin, orbital and lattice degrees of freedom~\cite{MMalvestuto_2011}. Their physical properties depend on the change of the number $n$ of the RuO$_{6}$ octahedra layers in the unit cell; and they range from the unconventional superconductivity in Sr$_{2}$RuO$_{4}$ ($n=1$)~\cite{Ishida_1998,AMackenzie_2003}, the quantum critical metamagnetism and nematicity in  Sr$_{3}$Ru$_{2}$O$_{7}$ $(n=2)$~\cite{SAGrigera_2001,RABorzi_2007,RSPerry_2001}, anisotropic ferromagnetism and proposed orbital-dependent metamagnetism in  Sr$_{4}$Ru$_{3}$O$_{10}$  (n=3)~\cite{YJJo_2007,MKCrawford_2002,GCao_2003,ZQMao_2006} and the spontaneous itinerant ferromagnetism in SrRuO$_3\,(n=\infty)$~\cite{LKlein_1996}. On the device applications side, members of RP-structured strontium ruthenates have been integrated in oxide electronic devices due to their good stability and structural compatibility with other correlated oxides. For example,  SrRuO$_3$ has been widely used as a conductive electrode in diverse oxide devices and heterostructures/superlattices~\cite{GKoster_2012}.  SrRuO$_3$/Sr$_{2}$RuO$_{4}$ hybrid structures have been explored in tunnel junction devices to establish superconducting spintronics utilizing ferromagnet/superconductor heterostructures~\cite{MSAnwar_2016}. Also, the Sr$_{3}$Ru$_{2}$O$_{7}$/SrRuO$_3$ and Sr$_{4}$Ru$_{3}$O$_{10}$/SrRuO$_3$ heterostructures have been reported to grow epitaxially on conventional oxide substrates~\cite{WTian_2007}, which could pave a way for the precise exploration of the known size effects in the magnetic properties of these compounds~\cite{YLiu_2016} by using ultra-thin devices fabricated from such heterostructures. All these behaviours show that Sr$_{n+1}$Ru$_{n}$O$_{3n+1}$ materials provide a fascinating playground to explore both novel quantum phenomena and diverse oxide-based electronic device directions.
\paragraph*{}We focus on the three layered member of the strontium ruthenate series. The triple-layered, Sr$_{4}$Ru$_{3}$O$_{10}$, is a quasi-two-dimensional metal with an orthorhombic unit cell, which has attracted attention due to its unique magnetic properties that are marked by the coexistence of ferromagnetism and itinerant metamagnetism~\cite{YJJo_2007,DFobes_2010,MKCrawford_2002,GCao_2003}.
 For a field applied along the $c$ axis $(H\parallel c)$, Sr$_{4}$Ru$_{3}$O$_{10}$  displays a ferromagnetic transition with a Curie temperature of $T_{\text{Curie}}\approx 105$ K, followed by another sharp transition at $T_{\text{M}}\approx 60$ K that is accompanied by a resistive anomaly in transport measurements~\cite{MKCrawford_2002,GCao_2003,ZXu_2007,YJJo_2007,FWeickert_2017}.  
For temperatures below $T_{\text{M}}$, the application of a magnetic field within the $ab$ plane $(H\parallel ab)$ induces a metamagnetic transition (i.e., a superlinear increase in the magnetization) at about 2.5 T~\cite{ZQMao_2006,DFobes_2010}. In addition, double metamagnetic behaviour was recently reported in magnetization data of Sr$_{4}$Ru$_{3}$O$_{10}$ with the second metamagnetic transition at a critical field slightly larger than the main metamagnetic jump~\cite{ECarleschi_2014}. However, the physical mechanisms of the exotic magnetic properties of Sr$_{4}$Ru$_{3}$O$_{10}$ below $T_{\text{M}}$ still are not well understood and remain elusive. For example, early specific-heat data exhibited no anomaly around $T_{\text{M}}$, which indicates that this second transition may not be an actual thermodynamic phase transition~\cite{XNLin_2004}; whereas recent magnetoresistance measurements suggested that the second magnetic transition at $T_{\text{M}}$ originates from a spin reorientation~\cite{YLiu_2018}. The metamagnetic transition was suggested to be an orbital-dependent effect, originating from the 
 Ru${\,4d}$ magnetic moments, where the Ru${\,4d_{xz,yz}}$ orbital is responsible for the metamagnetic transition while the $4d_{xy}$ orbital is ferromagnetic in the ground state~\cite{DFobes_2010,YJJo_2007}. On the other hand, to explain the double metamagnetic transition, it has been proposed that the metamagnetic behaviours could also originate either from magnetic ordering of the two inequivalent Ru sites or from fine structure in the near Fermi level ($E_F$) electronic structure due to the presence of two van Hove singularities (vHS) in the density of states (DOS)~\cite{ECarleschi_2014}. These vHS would render the material magnetically unstable to the extent that a metamagnetic transition would be induced by the application of a magnetic field.

Although the magnetic properties of Sr$_{4}$Ru$_{3}$O$_{10}$ have been extensively investigated~\cite{YJJo_2007,YLiu_2016,MKCrawford_2002,GCao_2003,ZQMao_2006,DFobes_2010,ZXu_2007,FWeickert_2017,ECarleschi_2014,YLiu_2018,XNLin_2004,VGranata_2016,WSchottenhamel_2016,MZhu_2018,VGranata_2013,YLiu_2017_01,YLiu_2017,YLiu_2016,RGupta_2006,YNakajima_2009,MZhou_2005,FWeickert_2018} 
and a possible origin of magnetic fluctuations suggested to be the presence of flat bands near-$E_F$ in the electronic band dispersions, i.e., the vHS~\cite{ECarleschi_2014}, as already predicted~\cite{DJSingh_2001,IHase_1997} and experimentally confirmed~\cite{ATamai_2008,MPAllan_2013} for its parent compound Sr$_{3}$Ru$_{2}$O$_{7}$; little is know about the electronic band structure of Sr$_{4}$Ru$_{3}$O$_{10}$. Thus, a precise characterization of the electronic band structure of this material by means of experimental probes is highly desirable.  Angle-resolved photoemission spectroscopy (ARPES) is the best candidate because it is one of the most direct methods to measure the momentum-dependent electronic band dispersion of solids~\cite{ADamascelli_2004}. 
\paragraph*{}Here, we report the first synchrotron-based ARPES measurements on the near-$E_F$ band structure of Sr$_{4}$Ru$_{3}$O$_{10}$, including its Fermi surface (FS) topology. Firstly, we present experimental FS maps. From a careful and systematic analysis of the individual quasiparticle   bands that make up the FS maps, we are able to identify  four FS sheets in the first Brillouin zone (BZ) of Sr$_{4}$Ru$_{3}$O$_{10}$. Secondly, taking advantage of the wide spectral range with an intense and highly polarized continuous spectrum offered by synchrotron radiation facilities, we study the effect of changing different matrix elements on the electronic band dispersions and FS maps of Sr$_{4}$Ru$_{3}$O$_{10}$. Thirdly, we study the coupling of the electronic degrees of freedom with collective excitations arising from magnetic or phononic  modes. We resolve five kinks in the near-$E_F$ band dispersions with energies ranging from 30 to 69 meV. A detailed analysis of kink energies and their comparison with reported Raman and lattice dynamic calculations data suggest that the kinks originate from strong electron-phonon coupling present in this system. By referring to other members of the strontium ruthenate oxides of the RP series exhibiting similar kink structure, we infer that the electron-phonon correlations may be ubiquitous in Sr$_{n+1}$Ru$_{n}$O$_{3n+1}$ compounds. We finish with a discussion on flat bands close to the Fermi-level that could be the electronic origin of magnetic instabilities in this system and suggest that electronic structure calculations, as well as further ARPES measurements, are needed to explore further the link between the fine structure in the near-$E_F$ electronic structure and magnetic instabilities present in the Sr$_{4}$Ru$_{3}$O$_{10}$ material.
\section*{\Large{M\lowercase{ethods and} E\lowercase{xperimental} D\lowercase{etails}}}
\paragraph*{}Single crystals of Sr$_{4}$Ru$_{3}$O$_{10}$  were grown using the flux-feeding floating zone technique with Ru self-flux. Details on the growth procedure of the samples used in this study are reported in Refs.~\cite{RFittipaldi_2007,RFittipaldi_2011}. The structural quality, composition and magnetic as well as electrical properties of the single crystals were characterised by X-ray diffraction, energy dispersive microscopy, scanning electron microscopy, specific heat, resistivity and susceptibility measurements. Samples were found to be pure single crystals of  Sr$_{4}$Ru$_{3}$O$_{10}$. 
\paragraph*{}All ARPES data on Sr$_{4}$Ru$_{3}$O$_{10}$ presented here were taken at the beamline CASSIOP\'EE of the synchrotron facility SOLEIL using a Scienta R4000 electron energy analyzer, which allowed us to measure simultaneously many energy distribution curves (EDC) in an angular range of 30$^\circ$. Two single crystals of Sr$_{4}$Ru$_{3}$O$_{10}$, labeled as sample $A_1$ and sample $A_2$, of dimensions of approximately $1\times 1 \times 0.5 $ mm$^3$ were used during the experiment. Experimental results were found to be reproducible for these two samples. The samples were cleaved in situ at a pressure of $\sim 2 \times 10^{-10}$ mbar and kept at a constant temperature of 5 K throughout the experiment. The spectra presented here are constituted of high resolution $k$-space band dispersions and FS maps of Sr$_{4}$Ru$_{3}$O$_{10}$ taken at photon energies of 37 eV, 60 eV and 110 eV. The angular resolution was 0.1$^\circ$ and the nominal energy resolution was 15 meV. The polarization of the incoming light was exploited in order to perform polarization-dependent ARPES experiments. Samples were mounted on the sample holder in such a way that their square in-plane crystallographic axes were aligned with the horizontal and vertical polarization of the incoming photon beam. In this way, FS maps and high-resolution cuts are symmetric, and unrotated in the first BZ, thus facilitating the identification of different matrix elements on the measured FS maps and high-resolution cuts.
\section*{\Large{R\lowercase{esults and} D\lowercase{iscussions}}}

\section{\Large{F\lowercase{ermi} S\lowercase{urface} \lowercase{of} S\lowercase{r}$_{4}$R\lowercase{u}$_{3}$O$_{10}$}}
\begin{figure*}[!t]
  \centering
\includegraphics[width=0.96\textwidth]{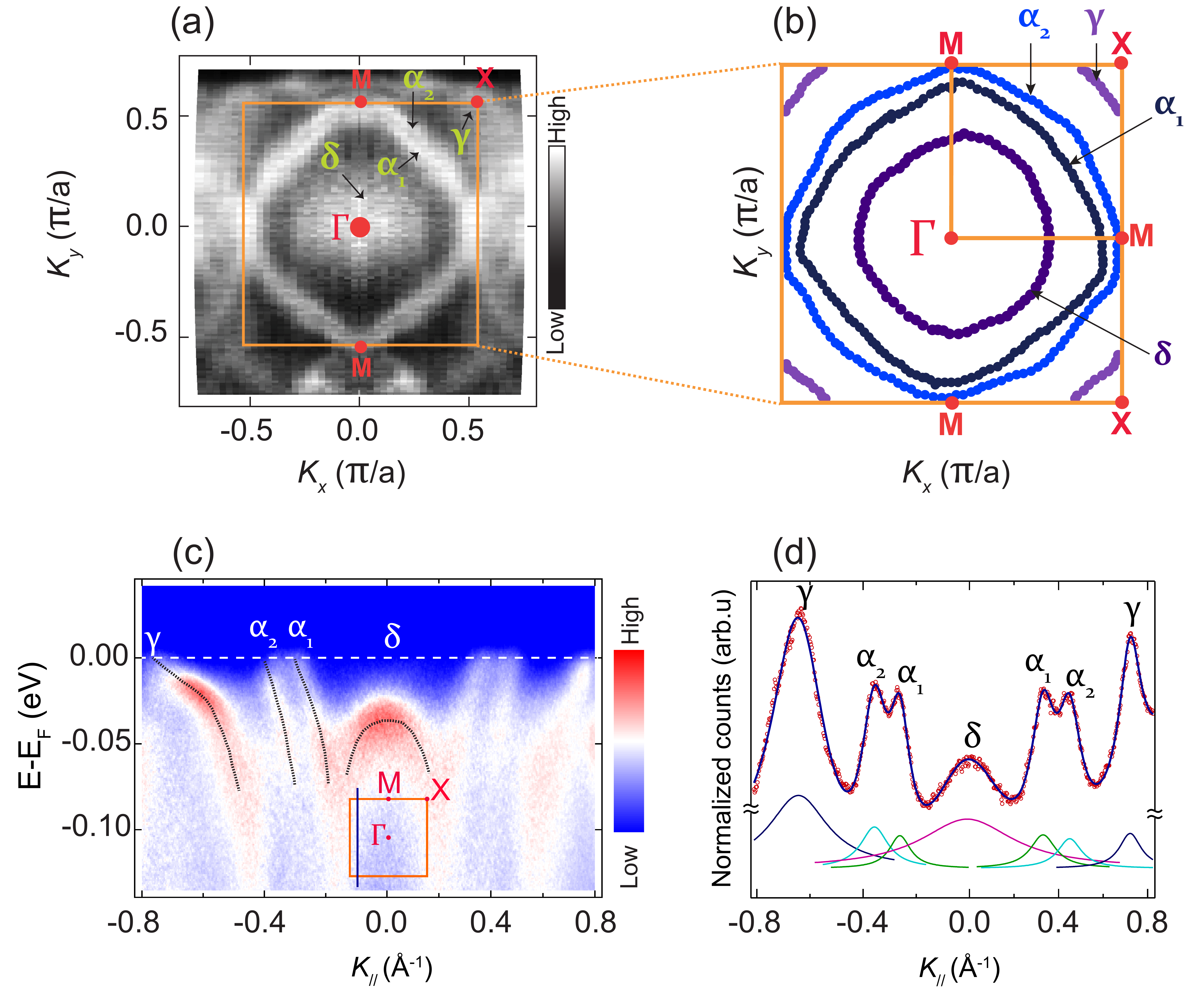}
  \caption{(a) FS map of Sr$_4$Ru$_3$O$_{10}$ for the sample $A_1$ measured using a photon energy of 60 eV in linear vertical polarization. The bright features are regions where the bands cross $E_F$.  The experimentally determined first BZ (orange square) and the positions of the high symmetry points are indicated. The FS map has been symmetrized with respect to the vertical line joining two $M$ points passing through the $\Gamma$ point. (b) Unsymmetrised contour map of the FS of Sr$_4$Ru$_3$O$_{10}$ extracted from the original FS map data. Four FS sheets $(\delta, \alpha_1, \alpha_2, \text{ and }\gamma)$ are identified in the contour map extracted from the analysis of individual cuts making up the FS map. The $\text{\textit{K}}_{_{//}}$  values in (a) and (b) are given in units of $\pi/a$, where $a$ is the in-plane crystalline constant of Sr$_4$Ru$_3$O$_{10}$. (c) A representative ARPES cut.  The blue vertical line in the inset orange BZ indicates the direction in which the cut was acquired. The black points overlaying the dispersing bands are extracted band dispersions for the E$_{\text{F}}$ crossing bands that give rise to the resolved FS sheets, which were clearly identified by fitting several MDCs. The slight intensity asymmetry visible around $\text{K}_{_{//}}=0$ could be ascribed to a misalignment of $\pm 1\degree$ during the mounting of the sample on the sample holder. (d) A representative MDC spectrum in red symbols together with the overall fitting (solid blue line) performed with Lorentzian line shapes. The individual Lorentzian peaks for the bands of interest are also shown.}
  \label{fig:Fig-01}
\end{figure*}

\paragraph*{} Figure~\ref{fig:Fig-01}\textcolor{blue}{a} depicts a FS map covering slightly more than the first BZ of Sr$_{4}$Ru$_{3}$O$_{10}$. A nominal energy resolution of $15$ meV was used and data have been integrated over an energy window of $\pm 12$  meV around $E_F$.  The bright features correspond to regions where the bands cross $E_F$. Since it is the first time FS maps of Sr$_{4}$Ru$_{3}$O$_{10}$ are reported, the resolved FS sheets were labelled keeping notations consistent with the previously resolved FS sheets of Sr$_{2}$RuO$_{4}$~\cite{ADamascelli_2000} and Sr$_{3}$Ru$_{2}$O$_{7}$~\cite{ATamai_2008}. 
For the assignment of the $M$ and $X$ points in the surface first BZ of Sr$_{4}$Ru$_{3}$O$_{10}$, we followed the notation consistent with the reported square BZ of Sr$_{2}$RuO$_{4}$~\cite{ADamascelli_2000,MWHaverkort_2008, ATamai_2019}. This is because our low temperature (5 K) ARPES data on the FS maps of Sr$_{4}$Ru$_{3}$O$_{10}$ suggest that its first BZ is not rotated as in the single layer system, Sr$_{2}$RuO$_{4}$; unlike the double layered system Sr$_{3}$Ru$_{2}$O$_{7}$, which has a rotated square BZ~\cite{ATamai_2008}.
 
\paragraph*{} We have resolved four FS sheets from the experimental ARPES measured FS map~[Figure~\ref{fig:Fig-01}\textcolor{blue}{a}]. The round contour centred at the $\Gamma$ point is identified as the $\delta$ FS sheet. A closer look at the map reveals three more FS sheets: two larger FS sheets centred around the $\Gamma$ point that are labelled $\alpha_1$ and $\alpha_2$, respectively; and a smaller one centred around the $X$ points, labelled $\gamma$. To illustrate the approximate shape of the resolved FS sheets, a contour map extracted from this map by fitting peaks from individual high resolution cuts making up the FS map is presented in Figure~\ref{fig:Fig-01}\textcolor{blue}{b}. The $E_F$  crossings of the four bands that give rise to these FS sheets have been clearly identified through the analysis of high resolution cuts taken at different $k$-directions in the BZ. Figure~\ref{fig:Fig-01}\textcolor{blue}{c} gives a representative ARPES cut taken in the direction parallel to the $\Gamma-M$ line in the first BZ. To follow the dispersion of these bands from $E_F$, several equally spaced momentum dispersive curves (MDCs) have been fitted with Lorentzian functions and the extracted peak positions are plotted on top of the corresponding band (black dashed lines) in the ARPES cut. Figure~\ref{fig:Fig-01}\textcolor{blue}{d} shows a representative MDC taken at $E_F$ and fitted with Lorentzian functions. From the MDC analyses, the slopes of the $\alpha_1, \alpha_2$ and $\gamma$ bands at $E_F$ are calculated to be 0.58 eV·\AA, 1.14 eV·\AA \, and 0.16 eV·\AA, respectively. From these slopes, the effective masses have been calculated for individual FS sheets using Fermi velocities determined along different band dispersions. A detailed analysis of these four $E_F$-crossing bands indicate that the $\gamma \text{ and } \delta$ bands have hole-like character, whereas the $\alpha$ bands have electron-like character [Table~\ref{table_S1} and Figure~\ref{fig:Fig-S2} in Supporting Information].

Considering a tetragonal crystal structure, a single trilayer of RuO$_6$ octahedra contains 3 Ru$^{4+}$ ions each contributing 4 conducting electrons distributed over the 3 nearly degenerate $t_{2g}$ levels~\cite{MKCrawford_2002,ECarleschi_2014}. In a first approximation; from the electron count, considering that there are small orthorhombic distortions and magnetic exchanges are small, it is expected to have 3 Ru\,$ 4d\,\,t_{2g}\times 3$ layers in the unit cell resulting in 9 electrons that would then give rise to nine FS sheets. However, from our ARPES data, we have resolved only 4 bands crossing the $E_F$ giving rise to four FS sheets. One possible reason for these undetected bands could be the inter-layer splitting mechanism in this system. As indicated above, in the consideration of small orthorhombic distortions and small magnetic exchanges, we are left with bonding, non-bonding and anti-bonding type of bands due to the layer degree of freedom. Since the $d_{xy}$ bands are not dispersive along the $z$-axis, they will be degenerate such that 3 out of the 9 bands will make one FS sheet with $xy$ character.  The remaining bands originating from the $(xz,yz)$ orbitals will form bonding, anti-bonding and non-bonding bands due to the out-of-plane hybridization with a splitting of the order of 150 meV~\cite{MMalvestuto_2011,EPavarini_2004}. Since the $(xz,yz)$ orthorhombic splitting is small (below the resolution limit), we would then end up with 3 bands (2-fold degenerate) with $(xz,yz)$ character and 1 band (3-fold degenerate) with $xy$ character; thus a total of 4 bands at $E_F$ giving rise to four FS sheets. An alternative reason for undetected bands could be the experimental energy resolution, nominally 15 meV. It could be that there are more bands very close to each other that contribute to the same ARPES peak which are not experimentally resolved. This argument is supported by previous ARPES data on Sr$_3$Ru$_2$O$_{7}$  in which an effective experimental resolution of $\sim 5.5$ meV was used  to resolve all the expected $E_F$-crossing bands~\cite{ATamai_2008}, suggesting that a complete determination of all $E_F$ crossing bands in Sr$_4$Ru$_3$O$_{10}$ is resolution limited. For more quantitative analysis and confirmation of these hypotheses with respect to the observed FS sheets, electronic structure calculations are indispensable, taking into account all the degrees of freedom present in this complex $4d$ system. 
\section{\Large{M\lowercase{atrix} E\lowercase{lement} E\lowercase{ffects} \lowercase{in} P\lowercase{hotoemission} S\lowercase{pectra} \lowercase{of} S\lowercase{r}$_{4}$R\lowercase{u}$_{3}$O$_{10}$}}
\begin{figure*}[!t]
  \centering
\includegraphics[width=1\textwidth]{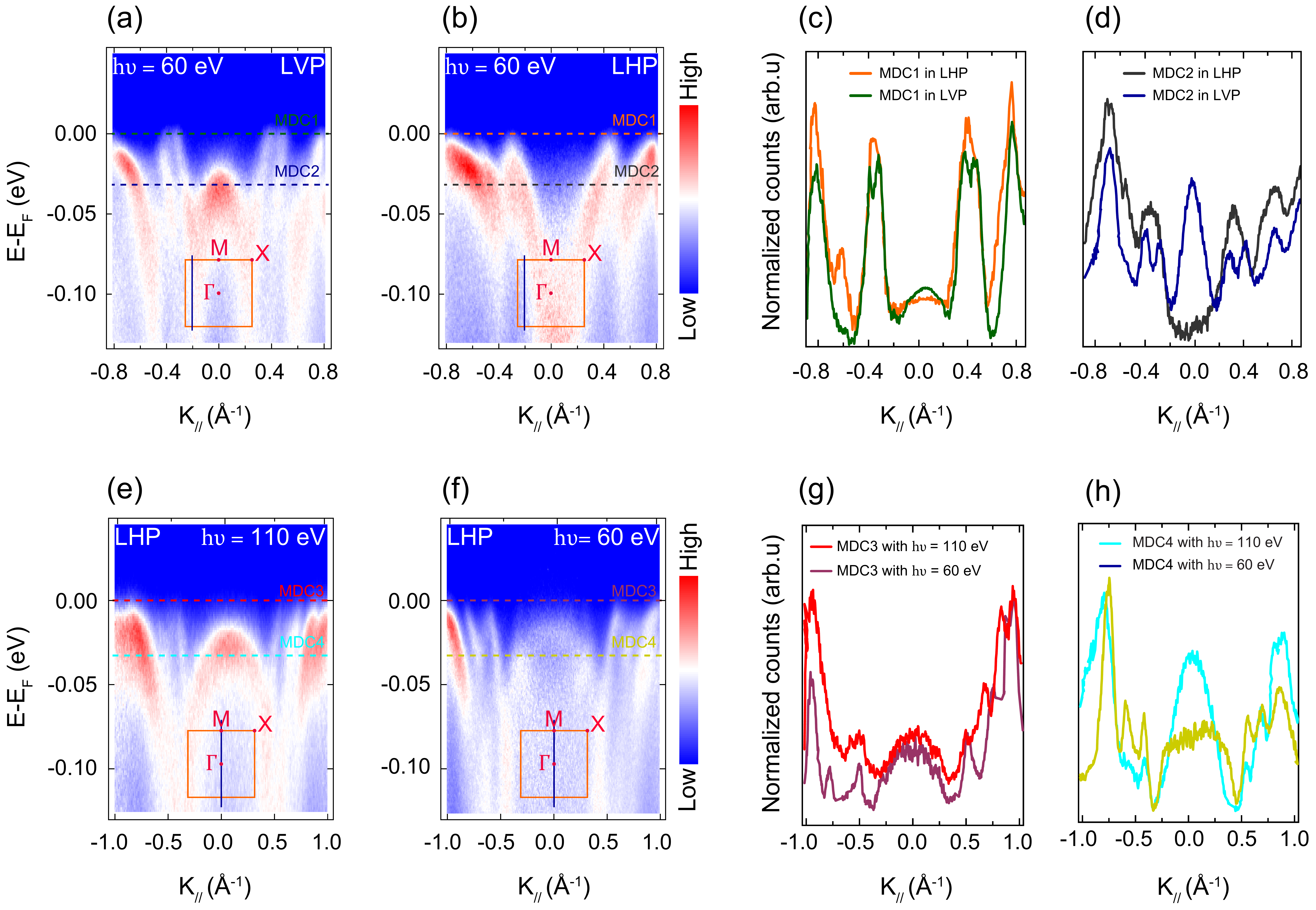}
  \caption{Photoemission matrix element effects on near-$E_F$ band dispersions of Sr$_4$Ru$_3$O$_{10}$ acquired at different locations in the BZ. Different features were resolved in ARPES spectra taken at the same location in the BZ with the same photon energy of 60 eV, but using different light polarizations (a) LVP (b) LHP; and with the same light polarization (LHP), but using different photon energies (e) 110 eV and (f) 60 eV. The blue vertical lines in the orange square BZ (insets) show the direction in which the cuts were acquired. The horizontal dashed lines indicate the energy positions (at $E_F$ and 30 meV) on which the MDCs in (c), (d) and (g), (h) were extracted. All spectra are from the sample $A_1$.}
  \label{fig:Fig-02}
\end{figure*}

\paragraph*{}To probe different features of the near-$E_F$ band dispersions, we have explored the effect of matrix elements on the quasiparticle band dispersions of Sr$_4$Ru$_3$O$_{10}$ using different photon energies (60 and 110 eV) and polarizations of the incident light, linear horizontal polarization (LHP) and linear vertical polarization (LVP). A detailed discussion on  photoemission matrix elements is presented in the Supporting
Information [Figure~\ref{fig:Fig-S3}\textcolor{blue}{a}-\textcolor{blue}{c}] and Refs.~\citep{ADamascelli_2003,SVBorisenko_2001,SMoser_2017}.
Exploiting the energy and polarization dependence of the ARPES matrix elements have already been demonstrated to be a powerful tool to extract the symmetry properties of the electronic states and disentangle the primary contributions of the main bands to the spectrum from those arising from secondary effects~\cite{MMulazzi_2006,MLindroos_2002,SVBorisenko_2001}.
\paragraph*{}  To exploit the effects of using different polarizations of the incident light, several ARPES cuts were taken using the same photon energy in the same position of the BZ, but with different light polarizations. Figure~\ref{fig:Fig-02}\textcolor{blue}{a} and \ref{fig:Fig-02}\textcolor{blue}{b} show representative ARPES band dispersions taken with the same photon energy of 60 eV in LVP and LHP, respectively. The band centred around the $\Gamma$ point is clearly resolved for the ARPES cut in LVP, while for the one in LHP the same band is not resolved at all; which is further confirmed by MDCs extracted at different locations on these ARPES cuts [Figure~\ref{fig:Fig-02}\textcolor{blue}{c} and \ref{fig:Fig-02}\textcolor{blue}{d}]. Moreover, for the ARPES cut taken in LVP, there are two clearly different bands, labelled $\alpha_1$ and $\alpha_2$ that are well resolved [Figure~\ref{fig:Fig-02}\textcolor{blue}{a}], whereas in LHP these two bands form one band. One can not differentiate their dispersions in the ARPES cut in LHP [Figure~\ref{fig:Fig-02}\textcolor{blue}{b}], as clearly shown by the extracted MDCs [Figure~\ref{fig:Fig-02}\textcolor{blue}{c} and \ref{fig:Fig-02}\textcolor{blue}{d}]. Fitting MDCs extracted from this ARPES cut at $E_F$, with Lorentzian functions [Figure \ref{fig:Fig-01}\textcolor{blue}{d}], the exact separation between the $\alpha_1$ and $\alpha_2$ bands is determined to be $\Delta K_{\parallel E_f}\simeq 0.102\pm 0.001 $ \AA. We have observed the same splitting for the FS maps acquired with incoming synchrotron light in LVP. Two FS sheets $\alpha_1$ and $\alpha_2$ that are centred around the $\Gamma$ point were resolved in LVP while only one FS sheet was resolved in LHP [see Figure~\ref{fig:Fig-S3}\textcolor{blue}{d} and \ref{fig:Fig-S3}\textcolor{blue}{e} in Supporting Information].
\paragraph*{} From the ARPES spectra acquired at different light polarizations, we have analysed the in-plane and out-of-plane band character of the Sr$_4$Ru$_3$O$_{10}$ system. When light is in LHP, it is possible to probe both in-plane bands and out-of-plane bands  by rotating the sample by a certain polar angle $\varphi$ around the vertical axis; while in LVP, one would only probe in-plane bands  [see details in Supporting
Information and Figure~\ref{fig:Fig-S3}\textcolor{blue}{c}]. From the two ARPES spectra shown in Figure~\ref{fig:Fig-02}\textcolor{blue}{a} and \ref{fig:Fig-02}\textcolor{blue}{b}, we speculate that the $\alpha_1$ and $\alpha_2$ bands have in-plane character ($d_{xy}, d_{x^2-y^2}$) since they are better resolved to be two separate 
bands when light is in LVP; while in LHP, they merge in one broader band. For the other resolved bands, the exploitation of light polarization together with experimental geometry showed no effect on their in-plane and out-of-plane characters; thus, band structure calculations of Sr$_4$Ru$_3$O$_{10}$ are therefore desirable to determine the character of all experimentally observed bands.
\paragraph*{}Furthermore, several ARPES cuts were taken using different photon energies, and their photoemission intensity was investigated. Figure~\ref{fig:Fig-02}\textcolor{blue}{e} and \ref{fig:Fig-02}\textcolor{blue}{f} show two ARPES cuts which were both acquired in LHP, but using different photon energies of 60 and 110 eV, respectively. These two energies were chosen as they showed the largest effects. Comparing these two ARPES band dispersions, we have observed that in general photoemission intensity is more enhanced for the ARPES cut taken with a photon energy of 110 eV. In particular, focusing on the band that gives rise to the $\delta$ FS sheet around the $\Gamma$ point, it is observed that for $h\nu = 110 $ eV the photoemission intensity is more enhanced and the band is better resolved. However, for  $h\nu = 60$ eV, the intensity of this same band is diminished and the band is not clearly resolved, as also demonstrated in the extracted MDCs in Figure~\ref{fig:Fig-02}\textcolor{blue}{g} and \ref{fig:Fig-02}\textcolor{blue}{h}. This observation shows the advantage of matrix elements to enhance the photoemission intensity of some band features while suppressing the intensity of others in electronic band dispersion.
\section{\Large{E\lowercase{lectron-phonon} I\lowercase{nteractions} \lowercase{in} B\lowercase{and} S\lowercase{tructure} \lowercase{of} S\lowercase{r}$_{4}$R\lowercase{u}$_{3}$O$_{10}$}}
\paragraph*{} The interaction of electrons with other excitations, e.g., phonons or spin excitations, often leads to 
anomalies in the energy-momentum dispersion relations that result in electronic band renormalization and sudden changes, referred to as \textit{kinks}, in the slope of the near-$E_F$ band dispersions~ \cite{AAKordyuk_2005,JFink_2007,TCuk_2004,ALanzara_2001,SAizaki_2012}. The kink energies are related to the relevant energy scales present in the investigated system, such as the coupling of electrons with other excitations. Kink structures have been reported in ARPES spectra at energies within 100 meV of $E_F$ in many correlated oxide systems, such as various cuprate superconductors~\cite{ALanzara_2001,XJZhou_2003},  SrVO$_3$~\cite{SAizaki_2012} and the RP-structured manganite La$_{2-2x}$Sr$_{1+2x}$Mn$_2$O$_7$~\cite{ZSun_2006}. Here, we study kink structures  in the electronic band dispersion of Sr$_4$Ru$_3$O$_{10}$.

\begin{figure*}[t]
  \centering
\includegraphics[width=1\textwidth]{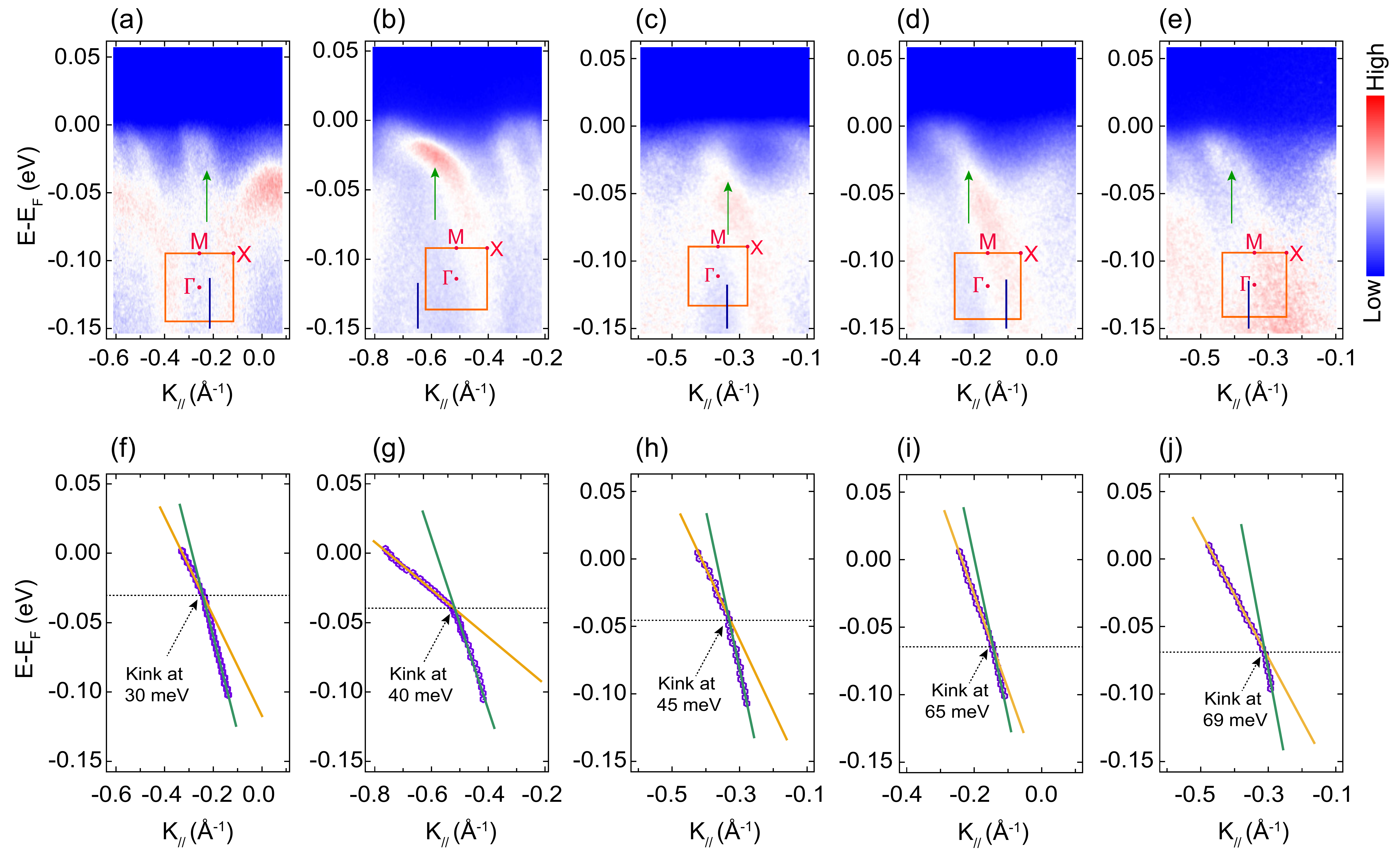}
  \caption{Kinks in the band dispersions of Sr$_4$Ru$_3$O$_{10}$ for the sample $A_1$.  (a)-(e) The photoemission intensity spectra  measured at different positions in the first Brillouin zone as shown in the inset of each cut. The blue vertical line in the BZ shows the direction in which the cuts were acquired. The green arrow in each cut indicates the band of interest. (f)-(j) Band dispersion (purple symbols) extracted from the fitting of MDCs extrapolated from the corresponding cuts in the upper panels for the band of interest. Black dashed arrows point to where kinks are observed, at the intersections of two linear fits to the low energy data. A possible second kink is visible in Figure 3(h) at approximately 19 meV. A  plausible origin for this kink is discussed in the final section of the Supporting Information.}
  \label{fig:Fig-03}
\end{figure*}

\paragraph*{}Figure~\ref{fig:Fig-03}\textcolor{blue}{a}-\textcolor{blue}{e} depict ARPES spectra of Sr$_4$Ru$_3$O$_{10}$ in selected locations of the BZ that were acquired using a photon energy of 60 eV in LVP. The bands of interest are indicated by green arrows in the top panels of the figure. These bands change slope as they approach $E_F$ and their photoemission intensity distribution varies. Some of the bands are 
more enhanced and better resolved [Figure~\ref{fig:Fig-03}\textcolor{blue}{b}] as they approach $E_F$ than others [Figure~\ref{fig:Fig-03}\textcolor{blue}{a}].  To quantitatively investigate the behaviour of these bands in the vicinity of $E_F$, several MDCs extracted from these spectra were fitted with Lorentzian functions and the binding energy of the MDC maxima, for the particular band of interest, have been followed down to binding energies of around $-0.1$ eV. Figure~\ref{fig:Fig-03}\textcolor{blue}{f}-\textcolor{blue}{j} show the extracted MDC peak maxima as a function of momentum. Kink features are clearly observed in the dispersion of these bands. To extract the kink energies, the near-$E_F$ low energy parts of the experimental dispersions have been fitted with two straight lines passing through the Fermi momentum $k_F$ [green and orange straight lines in Figure~\ref{fig:Fig-03}\textcolor{blue}{f}-\textcolor{blue}{j}]. Kink energies of 30, 40, 45, 65 and 69 meV were extracted. 
This energy range is comparable with what was previously reported in near-$E_F$ dispersions of other
Sr-based layered ruthenates~\cite{YAiura_2004,HIwasawa_2005,HIwasawa_2006,DEShai_2013,HYang_2016} and low energy kinks in the high-Tc superconductors~\cite{TCuk_2004,WMeevasana_2007}.
The presence of kinks in the electronic band dispersions of Sr$_4$Ru$_3$O$_{10}$ is further confirmed by the analysis of the MDC full width half maximum (FWHM) versus the binding energy of the peaks [Figure~\textcolor{blue}{S4}]. The FWHM of the MDC is directly proportional to the imaginary part of the self energy~\cite{AAKordyuk_2006}, which represents the scattering rate of the quasiparticles in the system. The scattering rate for this particular cut shows a discontinuity at an energy of $\sim 43$ meV, which is in very good agreement with the $40$ meV kink observed for this particular feature [Figure~\ref{fig:Fig-S4}\textcolor{blue}{c} in Supporting Information).

\paragraph*{}To understand the origin of kinks in the band structure of Sr$_4$Ru$_3$O$_{10}$, the extracted kink energies were compared to previous phonon modes from Raman spectroscopy and lattice dynamic calculations (LDCs)~\cite{MNIliev_2005, MNIliev_1999}. Remarkably, the kink energies have a good correspondence with the energy of the phononic modes from Raman spectroscopy and LDCs [Table~\ref{table_S2} in Supporting Information]. Thus, it is likely that the coupling of electrons to phonons plays a dominant role in the formation of kinks in Sr$_4$Ru$_3$O$_{10}$. As indicated above, the other three members of RP-structured ruthenates, Sr$_2$RuO$_4$, Sr$_3$Ru$_2$O$_7$ and SrRuO$_3$, also show kinks within the same energy scales which were proposed to originate from coupling between electrons and phonons in these compounds~\cite{YAiura_2004,HYang_2016}. In terms of crystal structure, these compounds belong to the layered perovskite oxide family, which are constituted of oxygen octahedra connected by oxygen atoms. The vibration of oxygen atoms, together with various tilts and distortions of similar oxygen octahedra leads to considerable phonon modes. When such phonon modes couple with electrons, renormalization effects [Figure~\ref{fig:Fig-S4}\textcolor{blue}{a}-\textcolor{blue}{b} in Supporting Information] with cut-off binding energies occur together with changes in the scattering rate near-$E_F$; thus leading to the formation of kinks in electronic band dispersions. Therefore, we suggest that electron-phonon interactions may be ubiquitous in Sr$_{n+1}$Ru$_{n}$O$_{3n+1}$ compounds impacting the physical properties of these compounds.

We would like to point out, on the other hand, that since the compound Sr$_4$Ru$_3$O$_{10}$ is a structurally distorted ferromagnet with RuO$_6$ octahedra rotations~\cite{MKCrawford_2002}, sensitive spin-phonon coupling is also expected in this compound. Notably, the 380 cm$^{-1}$ $B_{1g}$ phonon mode, reported previously to have a structural contribution to magnetic order in Sr$_4$Ru$_3$O$_{10}$~\cite{MNIliev_2005}, corresponds roughly to the ARPES kink of 45 meV. This suggests the presence of sensitive spin-lattice coupling and points to the possibility that some of the ARPES resolved kinks are a manifestation of coupling between structural and magnetic properties in this system.
\section{\Large{F\lowercase{lat }B\lowercase{ands }C\lowercase{lose to the} F\lowercase{ermi-level in} B\lowercase{and} S\lowercase{tructure} \lowercase{of} S\lowercase{r}$_{4}$R\lowercase{u}$_{3}$O$_{10}$}}
\begin{figure*}[!b]
  \centering
\includegraphics[width=1\textwidth]{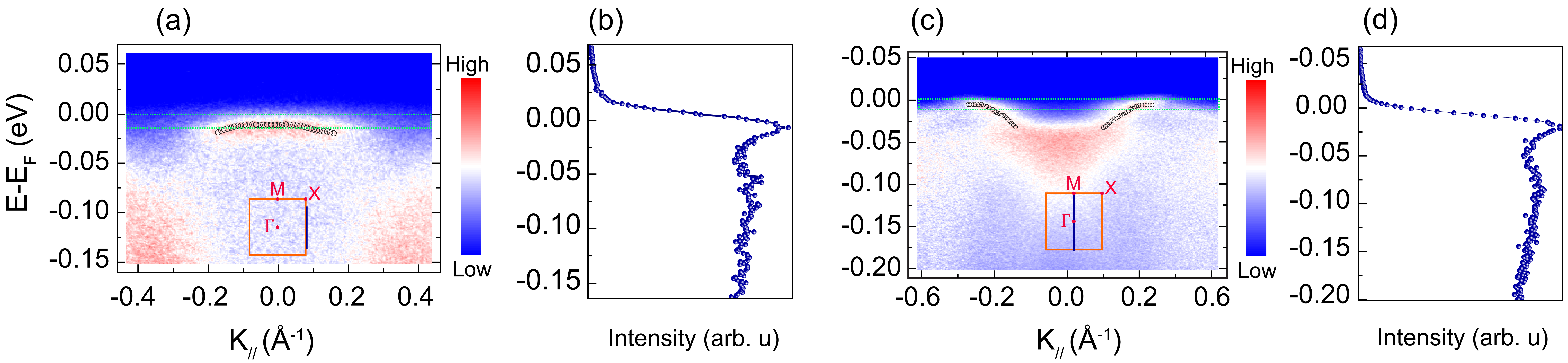}
        \caption{ARPES two dimensional cuts obtained in the proximity of $E_F$ measured using a photon energy of (a) 37 eV and (c) 60 eV for the sample $A_2$. The blue vertical line in the first BZ in the insert shows the direction in which the cuts were acquired. The black open symbol plotted on top of the flat bands of interest, (a) $\alpha_2$ and (c) $\delta$, correspond to the binding energy versus momentum quasiparticle peaks  obtained by fitting several EDCs taken from the $\gamma$ and $\delta$ bands with a Gaussian line shape. (b) and (d) Sharp peaks in the DOS located in a small energy window ($\lesssim8$ meV) around the Fermi-level, as indicated by light green dashed rectangles.}
     
  \label{fig:Fig-04}
\end{figure*} 
\paragraph*{}Finally, we reserve this last section to discuss the possible microscopic origin of the metamagnetic transition due to fine structure, within an energy range relevant for the Zeeman effect, present in the near-$E_F$ electronic structure of Sr$_4$Ru$_3$O$_{10}$.  It is an itinerant type of metamagnetism that is relevant in this system, which has also been reported in $3d$ and $f-$electron systems~\cite{TGoto_1997,NVBaranov_2009}. Itinerant metamagnetism is often predicted for electron systems with a two-dimensional DOS that has a logarithmically divergent vHS \cite{BBinz_2004,EPWohlfarth_1962,YTakahashi_1995}. The presence of such a vHS in the DOS renders the material magnetically unstable such that the application of a magnetic field induces a metamagnetic transition by shifting the vHS across $E_F$.  Previous band structure calculations and ARPES experiments have proposed this itinerant metamagnetism model to explain the complex band structure of the parent compound  Sr$_3$Ru$_2$O$_{7}$~\cite{ATamai_2008,DJSingh_2001,IHase_1997}. According to these works, it is the presence of such sharp peaks in the near-$E_F$ DOS (i.e., within a few meV from $E_F$) that are the electronic origin of magnetic fluctuations present in Sr$_3$Ru$_2$O$_{7}$~\cite{ATamai_2008,MPAllan_2013}. The natural cause of sharp peaks in the DOS is the presence of flat bands close to $E_F$. The same scenario could also be valid for Sr$_4$Ru$_3$O$_{10}$ given the close similarity between these two systems. However, we point out that our ARPES data on near-$E_F$ flat bands within an energy range relevant for the Zeeman effect are preliminary, therefore the discussion presented below remains at a speculative level.

\paragraph*{}  We have identified in different ARPES spectra of Sr$_4$Ru$_3$O$_{10}$ two flat bands of Ru $4d$  that present a complex DOS within a few meV of $E_F$, which is the energy scale relevant for 
metamagnetism [Figure~\ref{fig:Fig-04}\textcolor{blue}{a} and \ref{fig:Fig-04}\textcolor{blue}{c}]. The $\alpha_2$ and $\delta$ bands 
were observed to have a large DOS just close to $E_F$, a situation that is susceptible to give rise to vHS in the DOS [Figure~\ref{fig:Fig-04}\textcolor{blue}{b} and \ref{fig:Fig-04}\textcolor{blue}{d}]. The 
dispersion of the quasiparticle peak of these bands are very narrow and confined within a narrow energy window ($\lesssim 8$ meV) around $E_F$. 
This motivates a speculative discussion on the electronic origin of the metamagnetism in Sr$_4$Ru$_3$O$_{10}$. It is possible that through the 
application of an external magnetic field, the $\alpha_2$ and $\delta$ FS sheets may be spin polarized at a field of $T_M\approx 2.5$ T, where $T_M$ is the magnetic field at which the metamagnetic transition in Sr$_4$Ru$_3$O$_{10}$ is observed the along $ab$-plane. The $\alpha_2$ and $\delta$ FS sheets could then jump discontinuously over $E_F$ and give rise to metamagnetic behaviour observed in this compound. 

\paragraph*{}  Nonetheless, it is important to emphasize the effect of resolution in order to resolve such vHS in the near-$E_F$ DOS. To resolve narrow bands in the band structure of Sr$_3$Ru$_2$O$_{7}$, A. Tamai \textit{et al}., used an ARPES experimental energy resolution of 5.5 meV. This was of great importance in their work as some carrier sheets in their data showed an occupied band width of only 5 meV~\cite{ATamai_2008}. An energy resolution greater than 10 meV or the presence of an impurity scattering contribution of this order to the linewidth of the resolved peaks would have rendered it impossible to resolve the narrow flat bands at $E_F$ and consequently not detect the vHS in the near-$E_F$ DOS. In our experiment on Sr$_4$Ru$_3$O$_{10}$, the FWHM of the sharpest peak, extracted through EDC fitting,  was $\sim 21$ meV. Thus, further ARPES experiments with a better energy resolution and band structure calculations are necessary to ascertain the role of these two flat bands, $\alpha_2$ and $\delta$, as potential candidates that could be responsible for the appearance of metamagnetism in Sr$_4$Ru$_3$O$_{10}$, as has already demonstrated in its parent compound~\cite{ATamai_2008,MPAllan_2013}.
\section*{\Large{C\lowercase{onclusion}}} 
\paragraph*{}  In summary, we have systematically investigated the electronic band structure of Sr$_4$Ru$_3$O$_{10}$ using synchrotron based-ARPES. This study has provided the first information on the near-$E_F$ band dispersions and FS of Sr$_{4}$Ru$_3$O$_{10}$ and the effect of changing different matrix elements on electronic band dispersions, as well as electronic correlation effects present in this compound.

A detailed analysis of different ARPES cuts showed that four bands are crossing the Fermi-level. These bands give rise to four FS sheets and the careful analysis of each FS sheet showed that $\delta$ and $ \gamma$  are hole-like while the $\alpha_1$ and $\alpha_2$ are electron-like FS sheets. Exploiting different matrix elements, by using different photon energies and changing light polarization, it was possible to better discern some features in band dispersions and FS maps of Sr$_4$Ru$_3$O$_{10}$. For example, using different light polarizations (LHP and LHP), it was possible to resolve the separation of $\alpha_1$ and $\alpha_2$ dispersing states in LVP, while in LHP these bands seemed to  form one single band. Five different kink features of energies ranging from 30 to 69 meV were resolved. These kink structures were found to be approximately equal to the energies of the vibrational states detected by Raman spectroscopy and lattice dynamic calculations studies. This observation, together with comparison to kink structures reported in other three members of RP-structured ruthenates, demonstrate that the kinks in Sr$_4$Ru$_3$O$_{10}$ originate from the strong electron-phonon coupling. We have also discussed the possible ubiquity of electron-phonon coupling in Sr$_{n+1}$Ru$_{n}$O$_{3n+1}$ systems. Finally, these ARPES data show two flat bands, $\alpha_2$ and $\delta$, that exhibit a complex DOS closer to the Fermi-level, suggesting that these two bands are the most likely bands that will give rise to vHS in the near Fermi-level DOS,
a situation which is favorable for magnetic instabilities in Sr$_4$Ru$_3$O$_{10}$ material. We like to finish by also pointing out that this work is inline with current efforts to have in the near future the first African synchrotron light-source built on the African continent~\cite{PNgabonziza_2019,SHConnell_2018,SHConnell_2019}, which is expected to boost research output for African scientists and their international colleagues conducting experimental research related to or enabled by synchrotron light-sources.
\newpage
\paragraph*{}
We acknowledge valuable discussions with Dr. Sergey Borisenko. P. Ngabonziza and B. P. Doyle thank Prof. Jochen Mannhart for his support and assistance in the writing phase of this manuscript. This work was financially supported by the South African National Research Foundation (NRF) through travel grant no. 74033. B. P. Doyle, P. Ngabonziza and E. Carleschi acknowledge financial assistance from both the Faculty of Science Research Council and the University Research Council of the University of Johannesburg. The authors would like to thank the synchrotron facility SOLEIL for provision of beamtime as user under the proposal no. 20100202.
\\
\\
\textbf{Author contributions}
P. N, E. C and B. P. D. conceived and designed the experiment. V. Z, E. C, B. P. D., P. N, A. T-I, and F. B. performed experiments at Synchrotron SOLEIL. R. F, V. G., M. C, and A. V grew and characterized the single crystals used in experiments. P. N, E. C, V. Z  and B. P. D. analyzed the data and wrote the manuscript with contributions from all co-authors. \\ \\
\textbf{Competing financial interests}: The authors declare no competing financial interests.\\
\\
\textbf{Data availability}: The data that support the findings of this study are available from P. N, E. C and B. P. D. upon reasonable request.
\newpage
\section*{\Large{R\lowercase{eferences}}}

\title{Fermi Surface and Electron-phonon Interactions in Band Structure of Sr$_{4}$Ru$_{3}$O$_{10}$ Probed by Synchrotron-based ARPES}

\onecolumngrid
\newpage
\setcounter{table}{0}
\setcounter{figure}{0}
\renewcommand{\thefigure}{S\arabic{figure}}%
\renewcommand{\thetable}{S\arabic{table}}%
\setcounter{equation}{0}
\renewcommand{\theequation}{S\arabic{equation}}%
\setstretch{1.5}

\begin{center}
\title*{\textbf{\Large{Supporting
Information:} \\ [0.25in] \large{{Fermi surface and kink structures in Sr$_{4}$Ru$_{3}$O$_{10}$ revealed by synchrotron-based ARPES}}}}
\end{center}

\begin{center}
\large{Prosper Ngabonziza$^{1,2}$, Emanuela Carleschi$^2$, Volodymyr Zabolotnyy$^3$,\\ Amina Taleb-Ibrahimi$^4$, François  Bertran$^4$, Rosalba Fittipaldi$^{5,6}$, Veronica\\ Granata$^{5,6}$, Mario Cuoco$^{5,6}$, Antonio Vecchione$^{5,6}$, Bryan Patrick Doyle$^2$}\\
\end{center}
\begin{center}
$^1$\small{Max Planck Institute for Solid State Research, D-70569 Stuttgart, Germany}\\
$^2$\small{Department of Physics, University of Johannesburg, P.O. Box 524 Auckland Park 2006, \\Johannesburg, South Africa}\\ 
$^3$\small{Physikalisches Institut, Julius-Maximilians-Universität Würzburg, \\Am Hubland,97074 Würzburg,Germany}\\
$^4$\small{Synchrotron SOLEIL, L’Orme des Merisiers, Saint-Aubin-BP48, 91192 Gif-sur-Yvette, France}\\
$^5$\small{CNR-SPIN Salerno, Via Giovanni Paolo II, 84084 Fisciano, Italy}\\ 
$^6$\small{Department of Physics, University of Salerno, Via Giovanni Paolo II, 84084 Fisciano, Italy}
\end{center}

\section*{\Large{D\lowercase{etermination of the }F\lowercase{ermi} L\lowercase{evel and}  e\lowercase{xperimental} e\lowercase{nergy} r\lowercase{esolution}}}
To determine the Fermi-level ($E_F$) for each sample measured and for each incoming photon energy used, several energy dispersive curves (EDCs) taken at different specific momenta were fitted with a Fermi-Dirac distribution function:
\begin{equation*}\label{four1}
f(E)=\frac{1}{1+\exp\left( \frac{E_{kin}-E_F}{K_B T} \right)};
\end{equation*}
added to a linear background. Here, $E_{kin}$ is the kinetic energy, $K_B$ the Boltzmann constant and $T$ the temperature. The linear background was introduced to take into account of the electrons produced by second order light coming from the undulator of the synchrotron and transmitted by the beamline optics.
\begin{figure*}[!t]
  \centering
\includegraphics[width=0.85\textwidth]{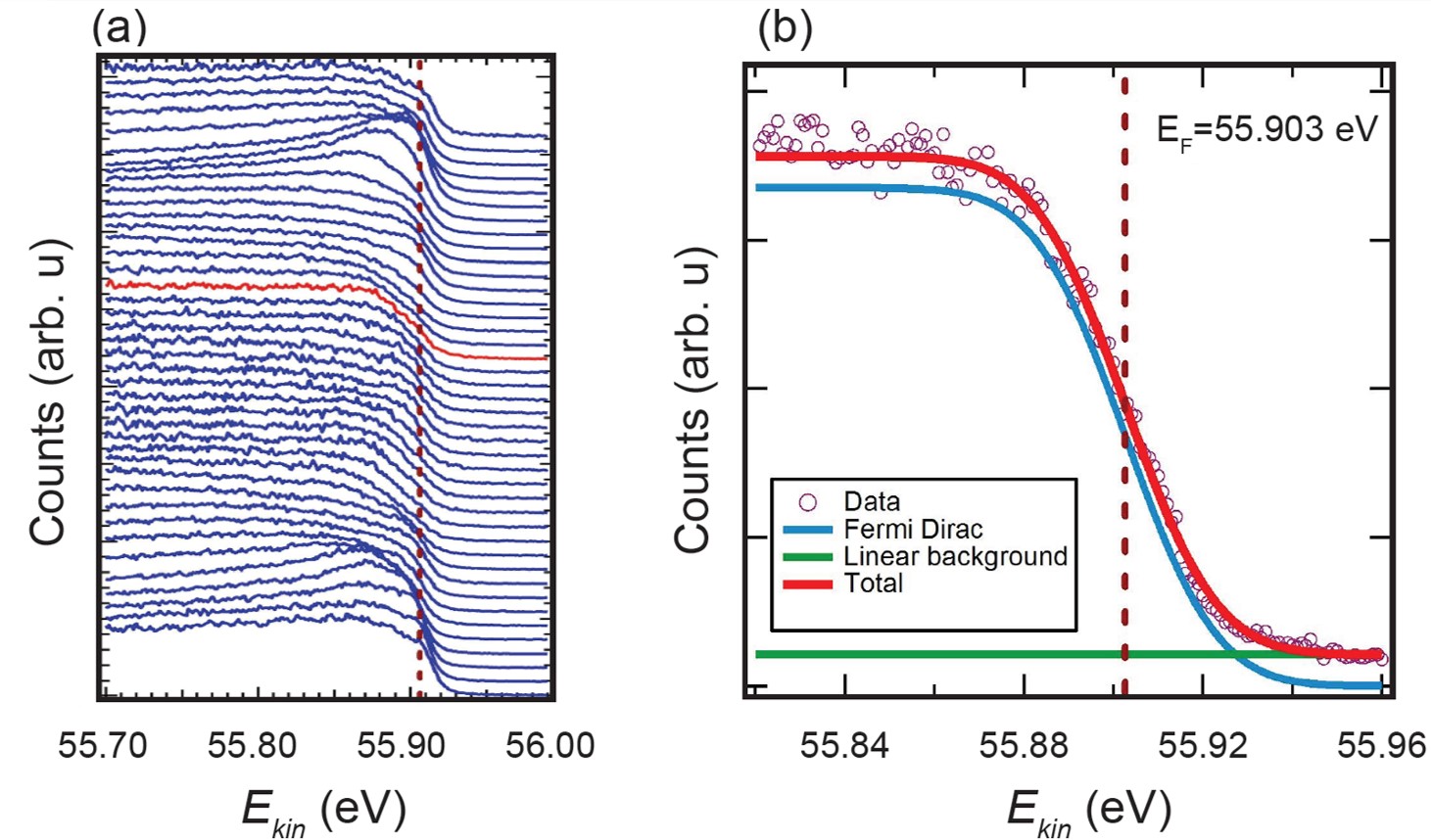}
  \caption{(a) Representative series of equally spaced energy dispersion curves (EDCs) extracted from an ARPES cut at different specific momenta for the sample $A_2$. The EDC spectrum in red is reported separately in purple open symbols (b) together with the overall fitting with a Fermi-Dirac distribution function added to a linear background to extract the Fermi energy. The blue curve is the Fermi-Dirac distribution function, the green line is the linear background while the red curve is their addition which is the fit to the data. }
  \label{fig:Fig-S1}
\end{figure*}
Figure \ref{fig:Fig-S1}\textcolor{blue}{a} shows equally spaced energy distribution curves taken at different momenta for an ARPES band dispersion acquired at a photon energy of 60 eV. To obtain the Fermi level $E_F$, a single EDC which has a Fermi-Dirac distribution like line-shape (the red curve) was taken from these equally spaced EDCs. Fitting this specific EDC [Figure~\ref{fig:Fig-S1}\textcolor{blue}{b}], the Fermi-level value of $E_F = 55.903$ eV is extracted. To check the accuracy of the obtained value of $E_F$, in each case 10 different EDCs have been fitted, and the change in the value of $E_F$ was found to be approximately $ \pm $  0.5 meV. The obtained $E_F$ was  used to set the binding energy scale in band dispersions and in energy distribution curves, using the expression $E_B=E_F-E_{kin}$. 
\section*{\Large{A\lowercase{nalysis of the} F\lowercase{ermi} S\lowercase{urface} S\lowercase{heets}}}
\begin{figure*}[t]
  \centering
\includegraphics[width=1\textwidth]{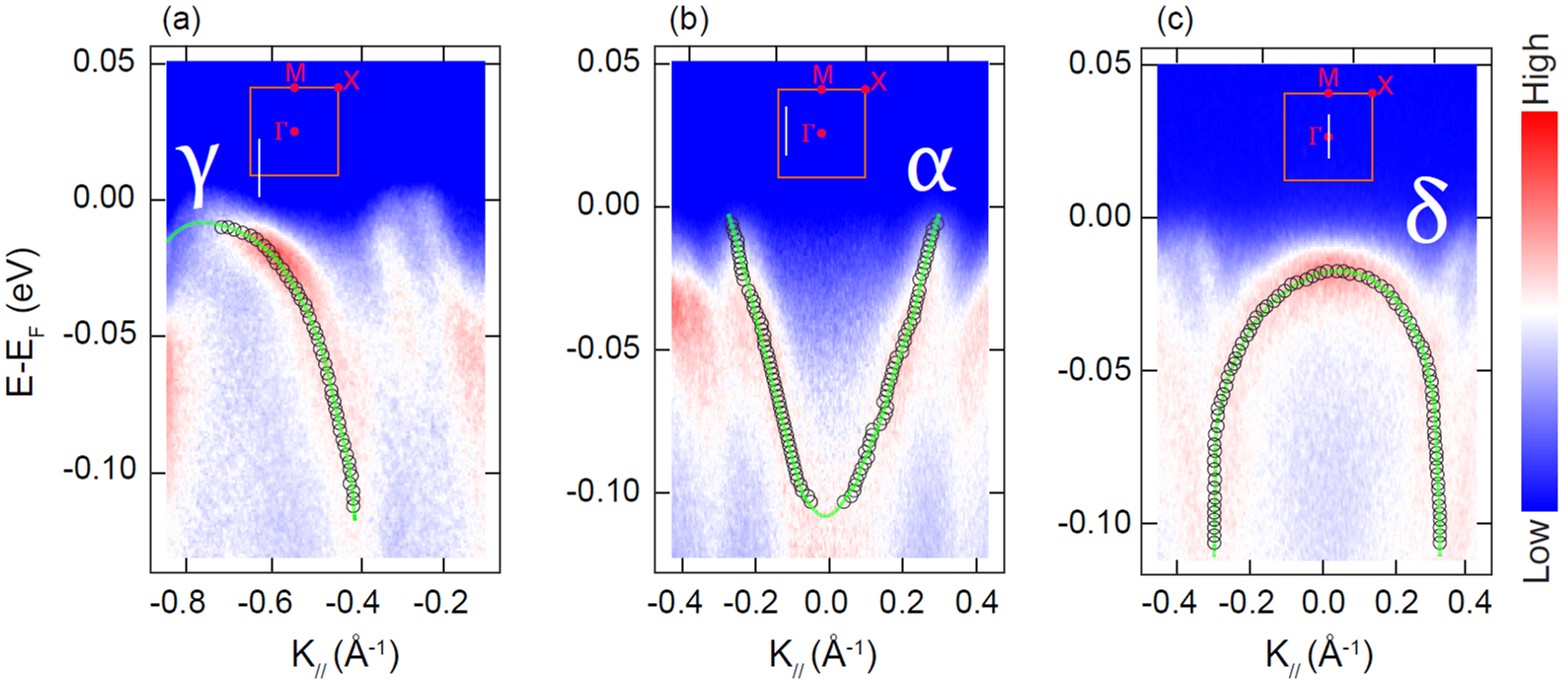}
  \caption{Investigation of electron and hole character of the ARPES resolved FS sheets of Sr$_4$Ru$_3$O$_{10}$ for the (a) $\gamma$ band, (b) $\alpha$ band, taken under conditions where the $\alpha_1$ and $\alpha_2$ seem like one band, and (c) $\delta$ band. The black open circles at the top of bands are MDCs derived dispersion of these bands. The dispersion of these bands show parabola-like behaviour (green dashed lines on top of the bands of interest). (Inset) The white vertical lines indicate the position and direction in which the cuts have been acquired with respect to the first BZ (square orange rectangle). The spectrum in (a) is from the sample $A_2$, whereas the spectra in (b) and (c) are from the sample $A_1$.}
  \label{fig:Fig-S2}
\end{figure*}
Analysis of several band dispersions, taken at different locations in the first BZ, allowed us to determine the electron or hole character of the four FS sheets. Figure~\ref{fig:Fig-S2}\textcolor{blue}{a}-\textcolor{blue}{c} show ARPES cuts measured with a photon energy of $60$ eV for the sample $A_1$. The black open circles on top of the bands of interest, $\gamma, \alpha$ and $\delta$ are MDC derived dispersions of these bands obtained by fitting several MDCs with Lorentzian functions. From these derived dispersions, it is clear that the $\gamma \text{ and } \delta$ bands have an inverted parabola-like dispersions as illustrated by the green parabola at the top of these bands; whereas the $\alpha$ bands have upward parabola-like behaviours. Thus, the $\gamma \text{ and } \delta$ bands will give rise to hole FS sheets whereas the  the $\alpha$ bands will give rise to electron FS sheets. Table~\ref{table_S1} summarizes  the extracted effective masses in the free electron approximation and the character (hole or electron).
\begin{table}[!t]
\caption{Characteristics of the FS sheets of Sr$_4$Ru$_3$O$_{10}$. The character of each pocket is indicated in square brackets, $h^+$ for hole pockets and $e^-$ for electron pockets. The effective masses of each FS sheet are also given. The effective mass of the $\delta$ FS sheet was not determined because the $\delta$ band is so broad. It was difficult to find its position at the Fermi-level and its $v_f$, but it gave a spectral weight at $E_F$.}
\begin{center}
\begin{tabular}{p{5cm}lp{2 cm}lp{2 cm}l p{ 2cm}l p{ 2cm}}
\hline
\hline
Character FS sheets && $\delta (h^+)$&&$\alpha_1 (e^-)$&&$\alpha_2 (e^-)$&& $\gamma (h^+)$ \\
\hline
&& && && \\
Effective mass && --&&$0.81$&&$0.41$&& $2.95$ \\
\hline
\hline
\end{tabular}
   \label{table_S1}
\end{center}
\end{table}
\newpage
\section*{\Large{P\lowercase{hotoemission} M\lowercase{atrix} E\lowercase{lements}}}
The expression of the measured photoemission intensity can be written as:
\begin{equation*}
I(k,\omega)=I_0(k,\omega,A) f(\omega)A(k,\omega);
\end{equation*}
where $f(\omega)$ and $A(k,\omega)$ are the Fermi-Dirac distribution function and the spectral function, respectively. To discuss the effect of the matrix elements on photoemission spectra, one looks at the function $I_0(k,\omega,A)$ which is proportional to the square of the matrix elements $\vert M^{\bold{k}}_{f,i}\vert^2$. It modulates intrinsic intensities according to the geometric experimental constraints, and it depends on photon energy, electron momentum and light polarization~\cite{ADamascelli_2004_SI}.
\begin{figure*}[!t]
  \centering
  \includegraphics[width=1\textwidth]{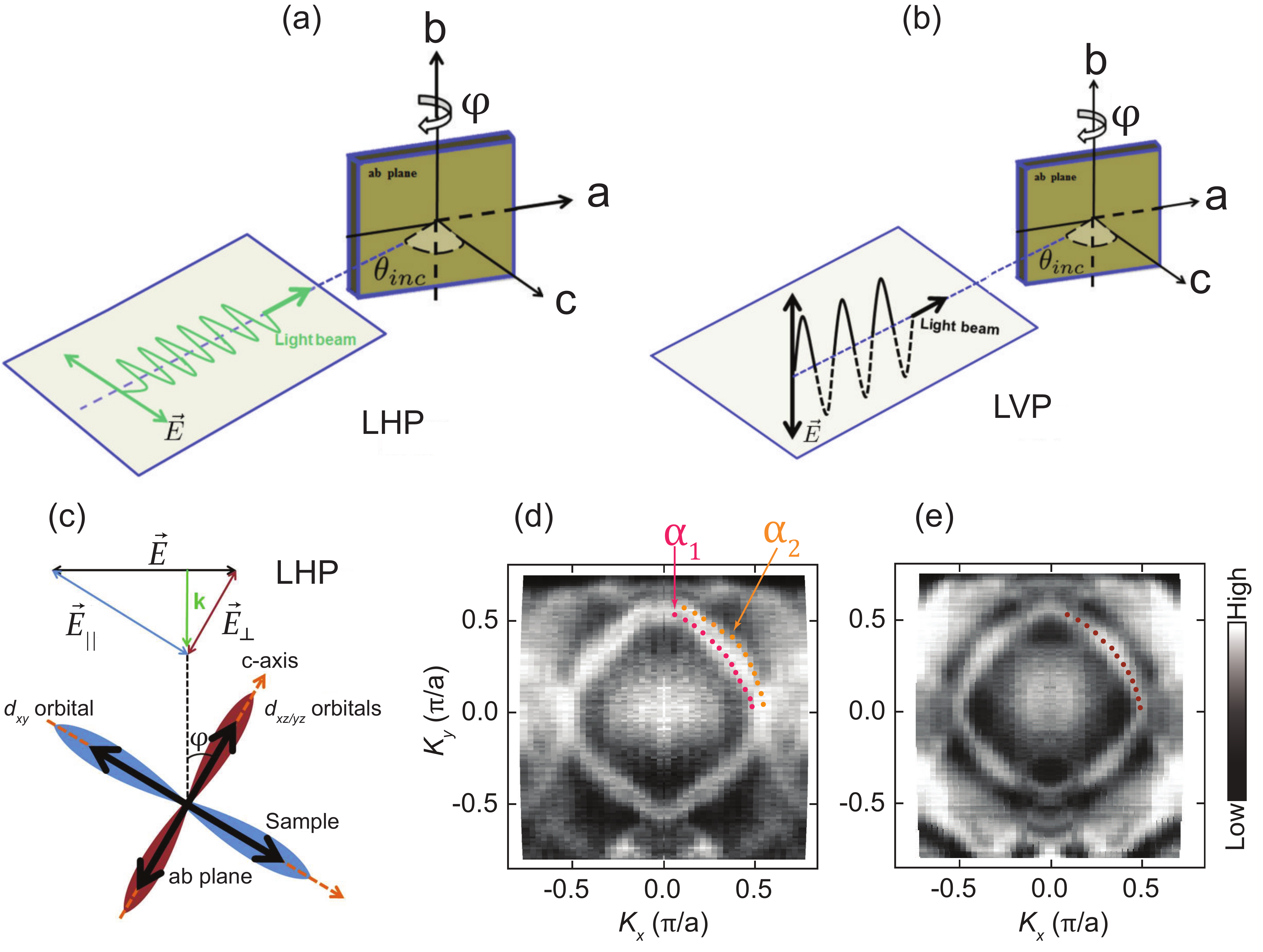}
  \caption{Matrix element effect using different light polarizations. Illustrative sketch showing the difference between light coming in (a) linear horizontal polarization (LHP) and (b) linear vertical polarization (LVP) with respect to the experimental geometry. (c) Schematic illustration of LHP light impinging on a sample with $d_{xy}$ and $d_{xz/yz}$ oriented orbitals. The parallel component of the electric field excites both in-plane $d_{xy}$ and out-of plane orbitals $d_{xz/yz}$, depending on the value of the polar angle $\varphi$. However, the perpendicular component of the electric field excites only in-plane orbitals $d_{xy}$. Fermi surface maps measured with a photon energy of 60 eV for the sample $A_1$ in (d) LVP and (e) in LHP. The markers plotted on top of the FS sheets are FS contours extracted from fitted MDC curves for the $\alpha_1$ and $\alpha_2$ bands. In LHP, the two bands cannot be distinguished.}
    \label{fig:Fig-S3}
\end{figure*}

During the ARPES experiment at the CASSIOP\'EE  beamline, spectra were acquired at different photon energies using two polarizations: linear horizontal polarization (LHP) and linear vertical polarization (LVP). This was done to take advantage of different matrix elements and also to determine the symmetry and character of the near-$E_F$ dispersing states. Figure \ref{fig:Fig-S3}\textcolor{blue}{a}-\textcolor{blue}{b} show the polarization configurations exploited in the ARPES experiment with respect to the experimental geometry. The manipulator used at CASSIOP\'EE  only allows rotation of the sample around the vertical axis of the manipulator angle $ \varphi$. This is why pre-orientation of the sample on the sample holder was necessary. Changing the sample orientation by varying $\varphi$ within a specific light polarization, it possible to probe in-plane and out-of-plane bands. When the light is in LHP [Figure \ref{fig:Fig-S3}\textcolor{blue}{a}], the electric field $\vec{E}$ is oscillating in the same plane as the orbit of the electrons in the ring and lies in the plane determined by the $a$ and the $c$ axes of the sample; while for light in LVP [Figure \ref{fig:Fig-S3}\textcolor{blue}{b}], the electric field is oscillating in the perpendicular plane to the orbit of the electrons and it is parallel to the $a$ axis. This implies that when light is in LVP, the electric field $\vec{E}$ is always in the $ab$ plane; while in LHP, the electric field $\vec{E}$ has two components: an out-of-plane component, given by $\vec{E}_{\perp}$, and an in-plane component, given by $\vec{E}_{\parallel}$. Thus, by rotating the sample with an angle $\varphi$ around the vertical axis, it is possible to probe both in-plane bands (for $\varphi \simeq 0 $) and out-of-plane bands (for higher values of $\varphi$) when incoming light is in LHP. However, with LVP, it is only possible to probe in-plane bands. This is evidenced in a more clear way by the sketch shown in Figure \ref{fig:Fig-S3}\textcolor{blue}{c}. For simplicity, we consider a general sample with $d_{xy}$ and $d_{xz/yz}$ orbitals that is illuminated by LHP light. In this configuration with light in LHP, it is possible to probe both in-plane band character ($d_{xy}, d_{x^2-y^2}$) and out-of-plane band character ($d_{xz}, d_{yz} $) by changing the polar angle $\varphi$. However, for in-coming light in LVP, it is only 
possible to probe in-plane bands ($d_{xy}, d_{x^2-y^2}$) as rotating the sample by an angle $\varphi$ around the vertical axis will yield no information on the near-$E_F$ out-of-plane dispersing states. Polarization dependence of incoming light was exploited in the ARPES experiment to probe symmetry properties of the electronic states of Sr$_4$Ru$_3$O$_{10}$. Figure~\ref{fig:Fig-S3}\textcolor{blue}{d}-\textcolor{blue}{e} show two FS maps taken using different polarizations of incoming light. For the FS map in LVP [Figure~~\ref{fig:Fig-S3}\textcolor{blue}{d}], two FS sheets ($\alpha_1$ and $\alpha_2$) are well resolved and they are clearly separate. In LHP, these two FS sheets are joined and form one FS sheet on the FS map [Figure~~\ref{fig:Fig-S3}\textcolor{blue}{e}].
\newpage
\section*{\Large{C\lowercase{orrelated} E\lowercase{ffect in the} B\lowercase{and} S\lowercase{tructure of}  S\lowercase{r}$_{4}$R\lowercase{u}$_{3}$O$_{10}$}}
\begin{figure*}[!h]
  \centering
  \includegraphics[width=1\textwidth]{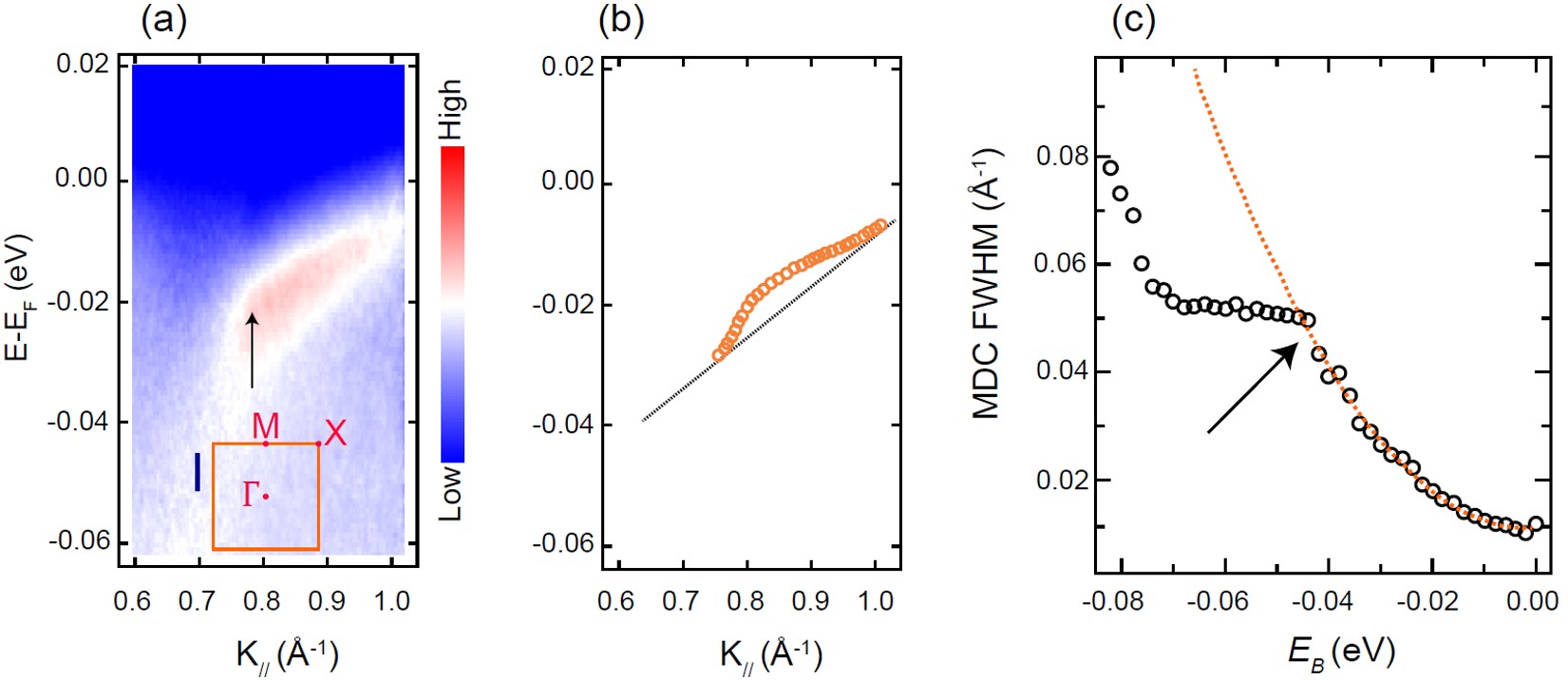}
  \caption{Renormalised bands in the electronic band dispersions of Sr$_4$Ru$_3$O$_{10}$ measured with a photon energy of 60 eV in LHP, in the direction parallel to the $\Gamma-M$ line for the sample $A_2$. The blue vertical lines indicate the position and direction in which the cuts have been acquired with respect to the first BZ (square orange rectangle in inset). The black arrows point to the band of interest. (b) The orange open circles are  peak maxima obtained by fitting MDCs extracted from the corresponding ARPES cut in (a). The black straight dashed-line represent band dispersions of a non-interacting system (the bare band dispersion). (c) A representative momentum (MDC) full width half maximum (FWHM) for the band of interest (the kink in Figure~\textcolor{blue}{3b} in the Main Text) plotted versus the binding energy of the peak. The dashed red curve represents a quadratic fit to the low-energy data. The arrow marks the position of the kink.}
    \label{fig:Fig-S4}
\end{figure*}

\begin{table}[!h]
\caption{ARPES kink energies from this study and previously reported phonon modes from Raman spectroscopy and lattice dynamic calculations (LDCs)~\cite{MNIliev_2005_SI}. The symmetry of the Raman modes, either $B_{1g}$ or $A_g$, is indicated. The comparison shows that the kink energies resolved from ARPES data are approximately equal to the energy of the phononic modes revealing a compatibility between their energy scales.\\}
\begin{center}
\begin{tabular}{C{4cm}lC{2 cm}lC{2 cm}l C{ 2cm}l C{ 2cm}C{ 2cm}C{ 2cm}}
\hline
\hline
ARPES && && && \\
(meV)&& 30 $ \pm \,3$&&40 $ \pm \,2$&& 45 $ \pm \,2$&&65 $ \pm \,1$ &69 $ \pm \, 4$ \\
\hline
&& && && && \\
Raman Spectroscopy && 29.15   && 38.08 && 47.1  && -- &72.95  \\ 
(meV)&& $A_g$&& $B_{1g}$&&$B_{1g}$&&--&$A_g$\\
\hline
&& && && &&  \\
LDCs && 28.66  && 40.5 && 43.92 &&63.52  &68.48  \\ 
(meV)&& $A_g$&&$B_{1g}$&& $B_{1g}$&&$A_g$&$A_g$\\
\hline
\hline
\end{tabular}
   \label{table_S2}
\end{center}
\end{table}
\newpage

\section*{\Large{R\lowercase{emarks on} e\lowercase{arly} s\lowercase{pecific} h\lowercase{eat}, s\lowercase{usceptibility and} r\lowercase{esistivity} m\lowercase{easurements for}  S\lowercase{r}$_{4}$R\lowercase{u}$_{3}$O$_{10}$}}

In the work by X. N. Lin \textit{et al}., the zero field specific heat experiment showed a weak anomaly at the Curie temperature while no anomaly was observed at the metamagnetic transition temperature~\cite{XNLin_2004_SI}. The magnetic field dependence of the specific heat revealed a large variety of anomalies with growing specific heat by increasing the field and abrupt jumps at critical fields applied in the  $ab$-plane and along the $c$-axis~\cite{GCao_2007_SI}. 
In the work by Cao \textit{et al.}, they also found a strong anisotropy in the magnetotransport with non-standard temperature dependence highlighting a possible non-trivial role of critical magnetic fluctuations and electron correlations~\cite{GCao_2007_SI}. The first susceptibility data were reported by M.K. Crawford \textit{et al.}, confirming the ferromagnetic/metamagnetic nature of the magnetism in the Sr$_4$Ru$_3$O$_{10}$ system~\cite{MKCrawford_2002_SI}. The magnetic structure has been further refined by neutron scattering experiments with the aim to set out the structural and magnetic interrelation as well as to establish the magnetic and orbital pattern within the unit cell~\cite{FForte_2019_SI,VGranata_2016_SI,VGranata_2013_SI}. In addition, several authors have also studied the resistivity characteristics of Sr$_4$Ru$_3$O$_{10}$ single crystals. For example, Mao \textit{et al.}~\cite{ZQMao_2006_SI} and Fobes \textit{et al.}~\cite{DFobes_2007_SI} have investigated the resistivity as function of magnetic field and temperature. They observed unusually transport behaviours around the metamagnetic transition. Their results showed sharp steps in the in-plane resistivity $\rho_{ab}$ for downward field sweeps between 1.75 T and 2.5 T and a remarkable non-metallic temperature dependence within this transition regime.

\section*{\Large{P\lowercase{ossible} L\lowercase{ow} E\lowercase{nergy} K\lowercase{ink}}}

In laser-based ARPES data on the parent ruthenate compound Sr$_2$RuO$_4$, Akebi and co-workers report several kinks in the low energy band dispersion of this material~\cite{SAkebi_2019_SI}. One such kink at very low energy ($\sim$ 8 meV) is attributed to magnetic excitations in relation to a peak found by inelastic polarised neutron scattering at 10 to 15 meV in the DOS of the material~\cite{PSteffens_2019_SI}. 
It is not immediately clear that one can transfer such observations to Sr$_4$Ru$_3$O$_{10}$ since one deals with a substantially different electronic system having triple Ru-O layers in the unit cell and a ferromagnetic ground state, thus already in a broken symmetry phase if compared to the paramagnetic state of Sr$_2$RuO$_4$. Moreover, there are no available inelastic neutron scattering data for Sr$_4$Ru$_3$O$_{10}$ which can guide the identification of the low-energy electron-mode coupling on the surface in the context of magnetic excitations. 
On the other hand, one can argue that in the ferromagnetic state with a critical temperature of the order of 100 K, the characteristic excitation energy of the spin-waves could be about 8 to 10 meV. With respect to having inelastic modes due to incommensurate or ferromagnetic fluctuations as in  Sr$_2$RuO$_4$, it is therefore plausible that the lowest energy kink arises from the coupling between the electron-mode and the spin-wave excitations. This issue might be important to assess the nature of the magnetic ground state on the surface of the Sr$_4$Ru$_3$O$_{10}$  especially in view of the lack of observed spin-split bands. 
Considering a possible connection with the flat band in Figure 4 in the main text, while the energy scales are comparable, it is difficult to argue about the occurrence of an electron-magnon coupling so strong that is able to induce spin polarons that are substantially non-dispersive. It is more plausible to expect that the Coulomb interaction is relevant to drive a significant renormalization of the bandwidth and in addition to that the coupling with magnons can make a weak kink in the dispersion. For the sake of completeness, due to the intricate magneto-structural effects, one cannot exclude that magnons and phonons cooperate at that energy scale to increase the effective mass of the electron modes.
\section*{\Large{R\lowercase{eferences} }}

\end{document}